# SDN-controlled and Orchestrated OPSquare DCN Enabling Automatic Network Slicing with Differentiated QoS Provisioning

Xuwei Xue, Fu Wang, Fernando Agraz, Albert Pagès, Bitao Pan, Fulong Yan, Xiaotao Guo, Salvatore Spadaro, Nicola Calabretta

*Abstract*—Optical switching techniques have the potential to enable the optical data center network (DCN) interconnections providing high capacity and fast switching capabilities, overcoming thus the bandwidth and latency bottleneck of present electrical switch-based multi-tiered DCNs. The rapid growth of multi-tenant applications with heterogeneous traffic require specialized quality of service (QoS) in terms of packet loss and latency to the DCN infrastructure. Slicing the DCNs into dedicated pieces according to the deployed applications, differentiated QoS and high resource utilization can be provided. However, slicing the optical DCNs still needs to be investigated because the Software-defined Networking (SDN) technique is developed for the electrical networks, not fully supporting the properties of the optical network. Additionally, Network Slices (NS) need to be automatically provisioned and reconfigured, to provide flexible slice interconnections in support of the multi-tenant applications to be deployed.

In this work, we propose and experimentally assess the automatic and flexible NSs configurations of optical OPSquare DCN controlled and orchestrated by an extended SDN control plane for multi-tenant applications with differentiated QoS provisioning. Optical Flow Control (OFC) protocol has been developed to prevent packet losses at switch sides caused by packet contentions. The extended OpenFlow (OF) protocol of SDN is deployed as well in support of the optical switching characteristics. Based on the collected resource topology of data plane, the optical network slices can be dynamically provisioned and automatically reconfigured by the SDN control plane. Meanwhile, experimental results validate that the priority assignment of application flows supplies dynamic QoS performance to various slices running applications with specific requirements in terms of packet loss and transmission latency. In addition, the capability of exposing traffic statistics information of data plane to SDN control plane enables the implementation of load balancing algorithms further improving the network performance with high QoS. No packet loss and less than 4.8 μs server-to-server latency can be guaranteed for the sliced network with highest priority at a load of 0.5.

*Index Terms*— Data center network; Optical interconnects; Optical switches; Flow control; Software defined networking.

## I. INTRODUCTION

THE escalation of traffic-boosting services and applications, such as cloud computing, Internet of Things (IoT), 5G mobile communications and high definition multimedia streaming, have significantly increased the traffic volume within the data centers (DC) [1, 2, 3]. The DCNs have to scale out to accommodate the tremendous increase in amounts of traffic [4, 5]. Besides the growth of DCN size, much more powerful servers are deployed to guarantee higher processing capabilities [6, 7]. The network scaling out and up requires the switching node handling tens of Tb/s aggregation traffic even in case of oversubscription. However, the implementation of high-bandwidth electrical switches is limited by the ASIC I/O bandwidth as the result of the scaling issue of ball grid array (BGA) package [8, 9]. Stacking several ASICs in a multi-tier structure could increase the switching bandwidth but at the expenses of extra latency and costly extra high capacity interconnections, leading to high cost and high power consumption switching architectures. Furthermore, the DCN scaling out/up impose enormous pressures to the network architectures in terms of capacity and latency. Present DCN architectures (Leaf-Spine, Fat-Tree, etc.,) with multi-tier switching layers are unable to provide the required network efficiency and flexibility to interconnect thousands of top of the racks (ToRs), each with tens-Tb/s aggregated traffic [10, 11, 12].

Benefitting from the data rate and format transparency, switching traffic in the optical domain overcoming the ASIC bandwidth bottleneck of electrical switches has been considerably discussed for the building of next-generation DCNs [13, 14]. Meanwhile, the high bandwidth of optical switches allows for flattening the network architecture then avoiding low capacity and large latency caused by hierarchical switching structures. Additionally, eliminating of power consuming O/E/O conversions at the optical switches significantly improves the energy-efficiency and cost-efficiency [15, 16, 17]. Recently, based on moderate-radix and

---

Manuscript received on June 1, 2019. This work is supported by Olympics project (ESTAR17207) and Metro-Haul project (G.A. 761727).

Xuwei Xue, Fu Wang, Bitao Pan, Fulong Yan, Xiaotao Guo and Nicola Calabretta are with the Electro-Optical Communications group, IPI research institute, Eindhoven University of Technology, 5612 AZ Eindhoven, The Netherlands (e-mail: x.xue.1@tue.nl).

Fernando Agraz, Albert Pagès, Salvatore Spadaro are with the Optical Communications group of Universitat Politècnica de Catalunya, Barcelona, Spain.

buffer-less fast optical cross-connect (OXC) switches, we proposed and numerically investigated OPSquare, an optical DCN architecture, where the parallel intra/inter-cluster switching networks feature with high capacity and low latency capabilities [18, 19].

Various applications and services managed by multiple tenants have to co-exist over the same optical network infrastructure for next-generation optical DCNs. The rapid growth of multi-tenant applications with various traffic flows impose their own set of heterogeneous requirements, in particular for packet loss and latency, to the network infrastructure, which requires dynamic QoS provisioning for each tenant [20, 21]. The over-provisioned and hierarchical full-electrical switching of present DC scenarios are not adapted to securely host mission-critical multitenancy applications. This calls out for developing the optical DCNs from the perspective of dynamic QoS supply. One promising strategy is to flexibly slice the infrastructures in a fully operable and manageable way to map the various QoS requirements for different tenants. Several research efforts have been proposed to provide differentiated QoS for optical DCNs [22, 23, 24, 25, 26]. In [22, 23], an optical flow concept so-called Express Path is established to enable the flows processing with various priority classes for solving the packet contention. However, the automatic reconfiguration, based on the network statistics, of the priority classes of the optical flow to provide a differentiated QoS has not been demonstrated. Similarly, the works presented in [24, 25] propose an SDN controlled slotted network operation with dynamic bandwidth allocation and differentiated QoS, but no automatic reconfiguration based on the monitored network statistics has been demonstrated. The network slicing with dynamic bandwidth allocation in [26] could potentially be automatically and flexibly reconfigured but also this has not yet been demonstrated.

An SDN control plane is introduced to manage the interconnections of OPSquare optical DCN [27]. Based on that, in this work, the Open Flow protocol of SDN in supporting of the optical switching characteristics is extended and deployed between the control plane and the data plane of the proposed OPSquare optical DCN. For this Open Flow integrations of OpendDayLight (ODL)/OpenStack (OS) control plane and OPSquare data plane, Open Flow agents are developed and deployed at the FPGA-based ToRs and switch controllers to collect and report the network resource topology and traffic statistics (number of lost and retransmitted packets) to the control plane. Based on this monitored network information, novel application engines have also been developed at the OpenDayLight/OpenStack platforms to automatically slice the network supplying the dynamic QoS for each network slice. Based on the extended Open Flow protocol, the OpenDayLight/OpenStack control plane with developed application engines and the proposed OPSquare optical DCN, we implement and experimentally assess for the first time the flexible and automatic network slices provisioning and reconfiguration of optical DCN in support of application-to-application connectivity with appropriate QoS provisioning for each tenant. Each NS consists of several virtual network functions (VNFs) hosted in servers locating in different racks. Meanwhile, the forwarding priority is automatically assigned to the traffic flows associated with each NS, according to the latency requirements of the applications. Once the virtual NSs is provisioned, the data plane switches the data traffic at nanoseconds scale benefitting from the fast reconfiguration of SOA-based switches, decoupled from the milliseconds of SDN operation time. In addition to this automatic NS provisioning, the NS reconfiguration has been implemented as well. Statistics of data plane (e.g. counts of retransmitted and lost packets, transmission latency) are continuously monitored and reported to the SDN control plane to trigger, if necessary, the proper NS reconfigurations. Additionally, benefiting from the multiple path connections among racks in the OPSquare architecture, the SDN controller can perform load balancing along with the NS reconfiguration to reduce packet losses, thus further enhance the required QoS supports.

The paper is organized as follows. Section II describes the SDN-controlled and orchestrated OPSquare DCN under investigation including the OPSquare infrastructure data plane, the specific control plane function as well as the OF extensions implemented to enable the communications between the SDN controller and the underlying OF Agents. The validation and assessment of the NSs provisioning and reconfiguration, differentiated QoS support via dynamic priority assignment, and automatic load balancing automatically triggered by the statistics collection of traffic flows are reported in section III. Finally, section IV concludes the paper by discussing the main results.

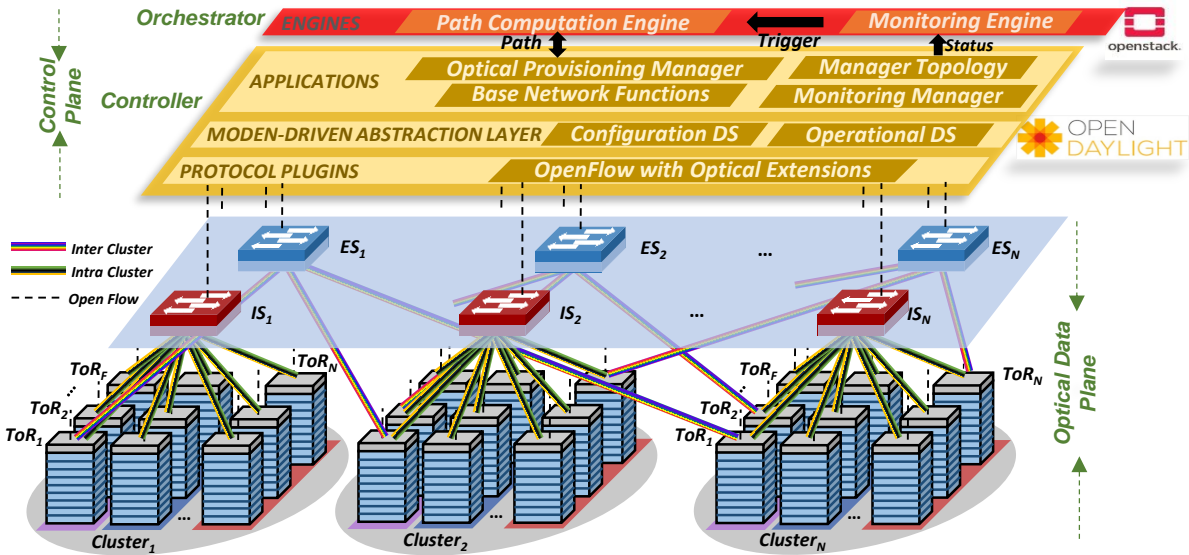

Fig. 1. SDN-controlled and orchestrated OPSquare DCN.

## II. SDN-CONTROLLED AND ORCHESTRATED OPSQUARE DCN

The overall architecture of the SDN-controlled and orchestrated OPSquare DCN is shown in Fig. 1. The data plane is organized following the principle of OPSquare architecture [18]. The OpenDayLight platform is set as the base SDN controller connecting the data plane by means of integrated OF Agents implementing an extended OF protocol. On top of the ODL controller, an OpenStack based Orchestrator is utilized to cooperate with the ODL controller to achieve the full automation of NS deployment featuring with dynamic flow priority assignment and automatic load balancing.

### A. OPSquare data plane

The OPSquare data plane comprising the distributed SOA-based optical switches and the grouped clusters is shown in Fig. 1[28, 29]. The data plane consists of N clusters and each of them groups K racks. K servers in each rack are interconnected via the FPGA-based ToR. Two bi-directional optical links are deployed at each ToR to access the parallel inter- and intra-cluster switching networks. As illustrated in Fig. 1, the N $N \times N$ inter-cluster optical switches (ESs) and N $N \times N$ intra-cluster optical switches (ISs) are dedicated for the traffic forwarding of inter- and intra-cluster, respectively. The $i$-th ToR of each cluster is interconnected by the $i$-th ES, where $i = 1, ..., N$. It is worth to notice that direct single-hop interconnection is supplied for ToRs locating in the same cluster, while at most two-hop interconnections are sufficient to link ToRs residing in different clusters. This way, the multiple path connections among racks in the OPSquare architecture enable the implementation of load balancing techniques to enhance the required QoS (lower packet losses) as well as increase the network fault tolerance.

The label channels are deployed to forward the label signals between the FPGA-based switch controller and ToRs (Fig. 2). The label signals indicate the destination ports of the corresponding optical data packets. Referencing to the forwarding look up tables (LUTs) stored in the controller, enable signals of switch gates are generated accordingly based on the received label signals to forward the optical packets to the destined ports. Packet contentions, i.e., optical packets from different ToRs destining to the same port of the optical switch, can happen due to the statistical multiplexing. In the data plane, the fast Optical Flow Control protocol between ToRs and IS/ES controllers is implemented on the label channels to prevent the packet loss at switch nodes caused by the contentions. The data packets are stored in the electrical buffer at ToRs waiting for the flow control signals coming from the switch controller nodes. The packets assigned with higher priority are directly forwarded by the switches while the packets with lower priority will be retransmitted in case of contentions. The flow control signals (ACK/NACK) generated by the IS/ES controllers are sent back to the ToRs for releasing the packets or requesting packet retransmission [30]. At the physical layer, the size of the electrical buffer at ToRs depends on the data traffic forwarding and retransmission time, network traffic load. Minimizing the overall buffer size will result in lower latency (i.e., shorter buffer queuing time) and lower cost and power consumption as well. However, small scale electrical buffer will cause packet loss when the buffer is over-filled especial for the network with a higher traffic load. Even if the Optical Flow Control protocol can prevent the packet loss at switch nodes, there still will be packet loss at the buffer sides when the network suffers heavy traffic or deploys smaller size buffer for network with critical latency requirement. The packet loss measured in this work is at the electrical buffer of each ToR due to the buffer overfill. Apart from the forwarding control of optical packets, the FPGA-based switch controllers report the counts of NACK signals indicating packet retransmissions to the SDN Controller which, in turn, forwards them to the Orchestrator for the trigger of load balancing to decrease the transmission latency and packet loss due to the buffer overfill. To synchronise the data packets and label packets arriving at optical switches/controllers at the same time, identical time is necessary for all the ToRs to arrange the delivery of these packets. In the proposed OPSquare network, the time of the switch controller at each cluster is distributed on the label channels to all the connected ToRs to unify the network time. Moreover, the fiber propagation delay of each label channel can be automatically measured to compensate for the distributed time. This guarantees the packets can be synchronised with less than 3.103 ns jitter. Benefitting from the modular feature of the OPSquare architecture, the network time distribution can be divided into cluster scale. At each cluster, the controller time is accurately distributed to all the ToRs on the independent label channels in a parallel way. Hence, the number of ToRs at each cluster does not deteriorate the performance of time distribution mechanism. The 3.103 ns packets arriving jitter can be implemented in the real large-scale network. With all these packets synchronised, label controlling mechanism and Optical Flow Control protocol can work accurately to forward the data packets to the correct destined ToRs.

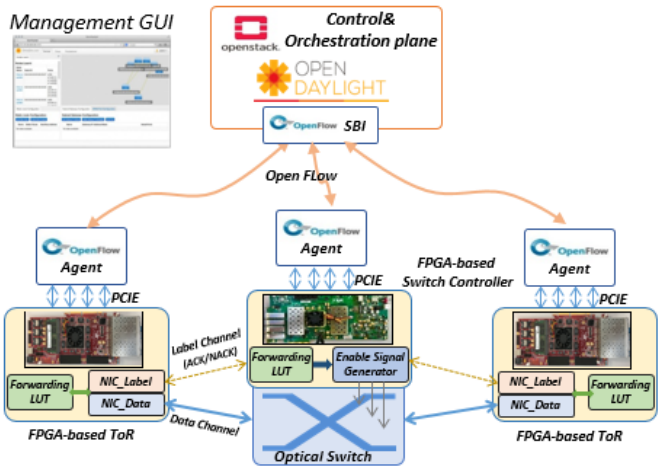

Fig. 2. Detailed interconnections of SDN-enabled OPSquare DCN. NIC: Network interface control.

### B. Extended Open Flow protocol for OPSquare network

The OF Agents deployed on top of FPGA-based ToRs and switch controllers are mediation entities between the control plane and data plane. Cooperating with OF protocol, these Agents enable the communications between the southbound interface (SBI) of ODL controller and the PCIe interfaces of data plane as shown in Fig. 2, whereby bridging monitor/report and configuration mechanisms at both sides. In particular, the Agents translate the OF commands generated from the ODL controller into a set of FPGA implemented actions through the proprietary interfaces and vice versa. Therefore, the resource topology and traffic statistics of underlying data plane can be monitored and reported to the control plane, while on the other hand, based on this monitored information, the control plane is able to automatically configure the OPSquare DCN by dynamically updating the forwarding LUTs.

The OF protocol deployed between the data and the control planes is leveraged to carry abstract monitoring information and network configuration commands. OF allows migrating the network control function out of the data plane devices to the control plane, since it enables the manipulation of the forwarding LUTs of such network devices. To this end, a group of dedicated attributes and messages are defined by the OF protocol to implement the separation of control and forwarding. Based on the OF FeatureReq – FeatureRep command pair, the resource topologies of data plane (i.e. number of switch ports, switching technology, etc.) are collected by the control plane.

TABLE I
OF PROTOCOL EXTENSIONS TO SUPPORT THE OPTICAL DCN

| Commands | Attributes | Extensions |
|---|---|---|
| FeatureRep | *ofb_capability* | Add fast optical switching features |
| FlowMod | *ofp_match* | Add optical flow and the used wavelength attributes as new matching fields |

Moreover, the ODL controller requests and collects the statistics of data plane via the OF command pair of StatsReq – StatsRep. The OF FlowMod commands conveying the operations (i.e. Flow creation, Flow modification or Flow deletion, etc) and flow characteristics (i.e. wavelength, load, etc.) are utilized to configure the connected hardware device and then the whole infrastructure. However, the present OF command specifications is customized for the electrical switches based DCN, so modifications and extensions are required to enable the OF commands in the proposed optical OPSquare DCN. In this paper, *ofb_capability* attribute indicating the OPSquare optical switching features is added in the FeatureRep command, whereby the ODL controller can recognize the devices with optical features. Moreover, the *ofp_match* fields are extended to configure the OPSquare data plane by means of the FlowMod command. More specifically, as illustrated in Table 1, the optical flow and the used wavelength attributes extend the *ofp_match* to match the destined network slice along with the ToR and/or switch input port. The regular OUPUT action is used to indicate the outgoing port of the ToR and/or switch The standard StatsReq – StatsRep commands cope with the collection of statistics of data plane (counts of retransmitted and lost packets), no extension necessary.

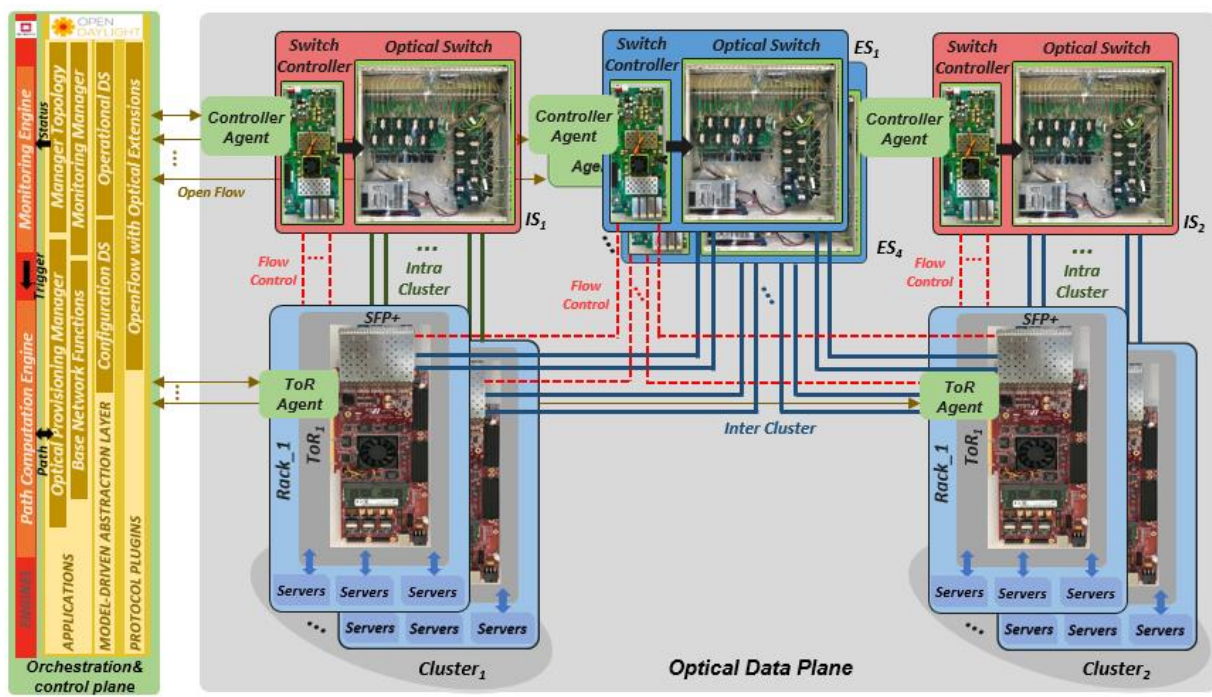

Fig. 3. Experimental set-up of SDN-controlled and orchestrated OPSquare DCN.

## C. Orchestration of the OPSquare DCN

To support the automatic network re-configurability and programmability, the ODL controller and an OpenStack-based Orchestrator are deployed on top of the data plane (Fig. 1). The ODL platform is set as the base SDN controller connecting the IS/ES controllers and ToRs by means of integrated OF Agents and extended OF protocol. The data plane layout and physical distribution information is stored in the Topology Manager (TM) as shown in Fig. 1. The Path Computation Engine (PCE) of the OpenStack-based Orchestrator, which relies on network topological information from the TM to provide the ODL controller with a ToR-to-ToR path computation service for NS deployment. Based on this calculated ToR-to-ToR path, the Optical Provisioning Manager (OPM) module sends the flow configuration (FlowMod) messages to the OF Agents, which configure the underlying devices (i.e., IS, ES and ToR) required to set up the specific network connectivity for NS provisioning and reconfiguration. Facilitated by the control plane and related protocols, once the NSs provisioned, the traffic flows generated by the running applications are automatically classified, recognized and associated to the given NS. Flow transmission priority can be dynamically allocated to each NS according to the QoS requirements of owning applications. Furthermore, the Monitoring Manager (MM) collects optical data plane statistics (counts of retransmitted and lost packets) and aggregates them into NS level metrics. Such aggregated information is collected by the Monitoring Engine (ME) of the Orchestrator to trigger the load balancing operations with benefits of the flexible path connections of OPSquare architecture whereby maintaining the expected low latency and packet loss.

## III. EXPERIMENTAL EVALUATION

Facilitated by the extended OF protocol and OF Agents, the Orchestration & Control plane enables the automatic operation of the connected OPSquare data plane. The automatic operations of control and data plane including NS provisioning and reconfiguration, priority allocation of application flows and load balancing according to the monitored statistics have been experimentally investigated.

The experimental set-up of the SDN-controlled and orchestrated OPSquare DCN is shown in Fig. 3. It consists of 8 FPGA-based ToRs equipped with OF Agents grouped into 2 clusters. Each FPGA-based ToR interconnects 4 servers at 10 Gb/s that generate Ethernet frames with variable and controllable load. The ToRs perform the statistical multiplexing of optical packets aggregating Ethernet frames with the same destination. The data packets are delivered by the ToRs via the 10 Gb/s intra-cluster data channel and inter-cluster data channel to the IS and ES, respectively. The optical switched in turn forward the packets to the destined switch ports according to the attached label signals carried by the label channels. 4 SOA-based ESs and 2 SOA-based ISs with corresponding FPGA-based switch controllers and OF Agents are utilized to interconnect all the racks. Meanwhile, the Optical Flow Control protocol is implemented between the ToRs and switch controller to prevent the packet loss. Furthermore, the ODL-based SDN controller is placed over the data plane to control the configuration of the infrastructure. The OS-based Orchestrator implements the automatic resources management operations.

## A. NS provisioning and reconfiguration

Fig. 4 shows the concept of NS allocation. Each NS comprises several virtual network functions (VNFs), which are deployed in the servers of the racks and interconnected through the optical DCN. The distribution of the VNFs and the interconnection is computed at the orchestration layer according to the functional requirements of the applications to be run and the availability of the data center infrastructure resources (i.e. computation and network). In the example depicted in the figure, based on the monitored data plane topology and resource, the PCE assigns NS1, NS2, NS3 and NS4 over the experimental infrastructure. In particular, $VNF_1$ and $VNF_2$ of $NS_1$ are assigned into Rack-1 and Rack-8, respectively. Afterwards, the Optical Provisioning Manager coordinates with the PCE to configure the best path among racks to provide connectivity for the VNFs. In this case, the selected paths between ToR1 and ToR8 for NS1 is ToR1↔ES1↔ToR5↔IS2↔ToR8. The LUTs of connecting nodes are accordingly initialized by the control plane. E.g., for the first hop of data packets generated from Rack-1 to destination Rack-8 at ES1, the port request of the label signal is 1 which indicates the destined output port of data packets is 1 connected with the ToR5. After the packets arriving at ToR5, a new port request of 2 generated automatically according to the stored LUT at ToR5 will be delivered to IS2 to control the second hop of data packets to output 2 which destines for the ToR8. Once the tenant owning NS1 needs to run a new application requiring more infrastructure resource (VNF). The SDN control plane monitors and collects these requirements from the data plane. Path Computation Engine cooperating with the Topology Manager allocates the VNFs residing in Rack-6 to the new NS1 (NS1'). Based on that, the Optical Provisioning Manager module sends the FlowMod commands to the corresponding OF Agents, then the Agents execute the reconfiguration instructions for the switch controllers (ES1, IS2) and ToRs (ToR1, ToR5, ToR6 and ToR8) involved in the new NS1 (NS1') to update the forwarding LUTs. It is worth to notice that, once the NSs provisioning is completed, application traffic is forwarded in the parallel intra-/inter-switching network, thus decoupling the data plane (switching at nanoseconds time scale) from the control plane (operating at milliseconds time scale). Moreover, exploiting the statistical multiplexing introduced by the OPSquare architecture, the servers, which host various VNFs, in the same rack can be allocated to different NSs. This optimizes the resources utilization of the optical data plane, leading to more application deployments.

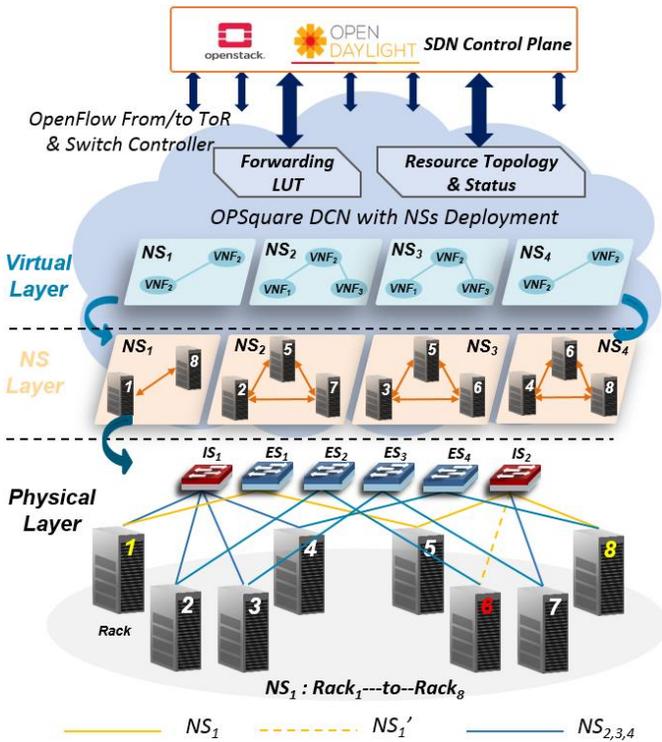

Fig. 4. NSs provisioning and reconfiguration.

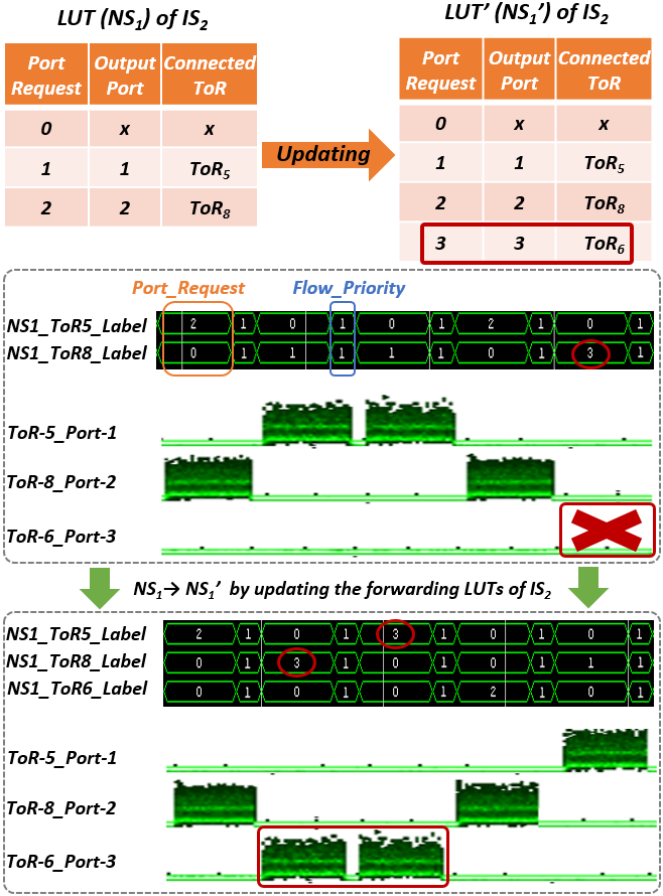

Fig. 5. Time traces for label signals and data packets before/after LUT updating (NS reconfiguration).

Fig. 5 illustrates the time traces of IS2 for label signals and data packets along with LUTs updating from the NS1 (LUT) to the reconfigured NS1' (LUT'). For the LUT of original NS1 at IS2, there are no entry routes to output Port-3 connecting with ToR6. Before the NS1 reconfiguration, data packets destined to ToR3 are blocked due to the absence of matching labels in the LUT. On the contrary, after the NS1 reconfiguration with the LUT updating, the data packets towards to ToR6 guided by the complemented label port request '3' are then correctly delivered. Note that, the transmission priority of traffic flows can be conveyed in the label signals as well. In this experimental case, the priority order of traffic flow is referenced as '1>2>3>4' where higher priority meaning lower transmission latency. We assume the application running in NS1 requires critical latency performance, whereby assigning priority '1' to all the traffic flow of NS1. As we can observe from Fig. 5, benefitting from this time distribution mechanism, the label signals of each time-slot arrive at the switch controller synchronously for the forwarding decision. Controlled by the switch controller, the data packets can be precisely forwarded to the destining ToRs as shown by the monitored optical traces.

The time components required for the reconfiguration of network slice is counted considering the whole workflows. The procedure has been repeated ten times and the average processing time is 125 ms consisting of the requirements collection of data plane (NS1 requiring more VNFs), algorithms processing at SDN control plane (Rack-6 to be allocated to NS1) and LUT updating of infrastructure (LUT' including Port-3). Moreover, the reconfiguration process does not affect other traffic flows, so data packets destined to ToR1 and ToR8 and other NSs perform hitless switching during the NS1 reconfiguration until the new LUT' updated.

### B. Dynamic priority assignment of application traffic flows

The proposed programmable OPSquare DCN features high resource utilization benefitting from the flexible NSs deployment. Nevertheless, as the traffic load increases, the traffic from different NSs may result in packet contentions at IS/ES nodes thus causing packet loss. The Optical Flow Control protocol introduced in section II has been implemented to prevent packet loss associated with contentions. However, the retransmission of blocked packets deteriorates the server-to-

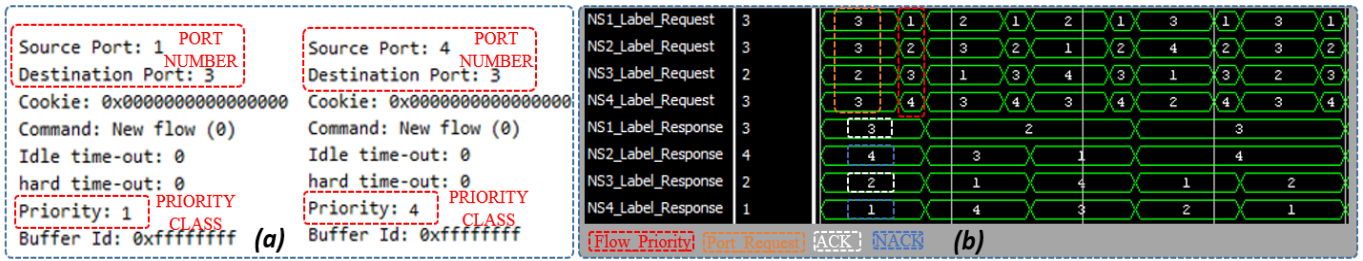

Fig. 6 (a) OF FlowMod commands to assign packet priority; (b) OFC signals monitored at the controller of IS2.

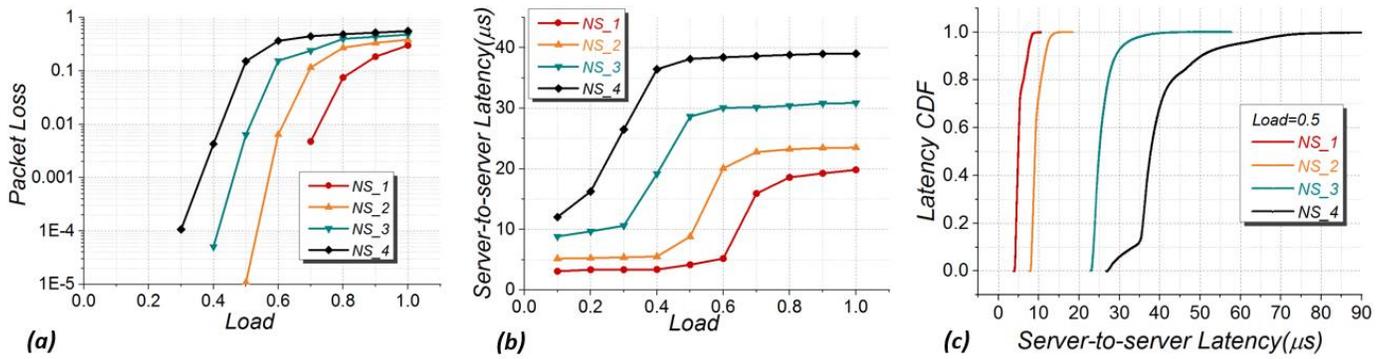

Fig. 7. (a) Packet loss (b) server-to-server latency versus traffic load for 4 NSs with specified priority; (c) Cumulative Distribution Function (CDF) of latency for 4 NSs at the load of 0.5.

server latency performance and furthermore, once the backup buffer is full at ToR, the incoming packets will be lost. To solve these issues, the transmission priority can be assigned to traffic flows during the NSs provisioning according to the QoS requirements of running applications. The flow with higher priority will be directly forwarded without retransmission thus avoiding transmission latency. In the SDN-enabled OPSquare architecture, the Orchestrator computes the flow priority of NSs based on the QoS requirements of the applications to be run in the slice. Upon the computed priority order, the Optical Provisioning Manager allocates the flow priority to each NS by the extended OF FlowMod commands as shown in Fig. 6(a). The proposed sliceable DCN provides 4 classes of priority to each NS. $NS_1$ hosting the latency-sensitive applications is assigned with the highest priority '1', while the lowest priority '4' is allocated to $NS_4$ where applications are latency insensitive. Following this rule, $NS_2$ and $NS_3$ are assigned with priority '2' and '3', respectively. The priority order of traffic flow is referenced as '1>2>3>4'. When contention happens, flows in $NS_1$ with highest priority will be forwarded directly. It is worth to note that packet retransmissions can occur in traffic flows of $NS_1$. In this case, conflicting packets of $NS_1$ will be forwarded following the Round-robin mechanism.

Fig. 6(b) illustrates the Optical Flow Control signals (label signals and ACK/NACK) from and to these 4 NSs monitored at FPGA-based controller of $IS_2$. $IS_2$ connects $ToR_5$, $ToR_6$, $ToR_7$ and $ToR_8$ in the cluster-2. Servers in these racks belong to $NS_1$, $NS_2$, $NS_3$ and $NS_4$, respectively. At every time-slot, the connected ToRs provide to the switch controller the label signals ($NS\_Label\_Request$) conveying the port destination and transmission priority of the associated data packets. Once packet contentions happen, the FPGA-based switch controller solves the contention according to the received port request and packets priority then sending ACK ($NS\_Label\_Response = NS\_Label\_Request$) and NACK ($NS\_Label\_Response \neq NS\_Label\_Request$) signals to the connected ToRs. As shown in Fig. 6(b), in case the $ToR_{5,7,8}$ send packets to $ToR_6$ through the switch output Port-3 of $IS_2$, $ToR_5$ of $NS_1$ (highest priority) receives an ACK signal to release the stored packet, while $ToR_7$ and $ToR_8$ of $NS_2$ and $NS_4$ get NACK signals to trigger the retransmission of the dropped packets. Thus, packets from $ToR_5$ are directly forwarded to the destined port avoiding extra retransmission latency.

Fig. 7 shows the packet loss and mean server-to-server latency for all the considered 4 NSs with specific flow priority versus different traffic load. Server-to-server latency is defined as the mean transmission latency from the source server to the destination server for all the optical links in the sliced network. Traffic load refers to Ethernet frames generated randomly between 64 and 1518 bytes by the connected servers. After 24 hours of stable network operation, the packet loss and latency of the experimental network as a function of the traffic load has been recorded. The recorded packet loss and latency are taken for 10 minutes for each traffic load value ranging from 0.1 to 1. More than 1 billion Ethernet frames are generated for the measurements at each rack within the 10 minutes measurement time. The experiment has been repeated 5 times and the mean packet loss and latency are calculated and shown in Fig. 7.The results confirm no packet loss for $NS_1$ up to load 0.7, while $NS_4$ with the lowest priority achieves the acceptable packet loss (0.04) for the application with lower QoS requirement at the load of 0.4. It is worth to notice that the packet loss of Fig. 7(a) is caused by the buffer over-fill under higher traffic load, not contributed by the packet loss at switch nodes, which has been prevented by the Optical Flow Control protocol. The retransmissions observed for the blocked packets of NS assigned with lower priority lead to an extra latency. On the contrary, the highest priority guarantees the mean latency of around 4.8 μs at the 0.5 traffic load and thus a high QoS for $NS_1$, which includes the 4.497 μs of buffer queuing time at the ToRs, 140 ns delay on the $2 \times 14$ m fiber as well as the FPGA processing time at TX and RX (163.5 ns). $NS_4$ presents the mean latency of 38.12 μs at high load of 0.5, which is compatible with applications with best-effort QoS requirement. We count the server-to-server latency for all the packets of the 4 NSs when the traffic load is 0.5. The Cumulative Distribution Function (CDF) of server-to-server latency is shown in Fig. 7(c). The results indicate that the server-to-server latency of $NS_1$ and $NS_2$ (the two higher priorities) has low variation with respect to the mean value of 4.8 μs and 9.7 μs, respectively. As a comparison, the latency distribution of $NS_4$ with the lowest priority is more dispersed, with latency distribution from 26 μs to 90 μs.

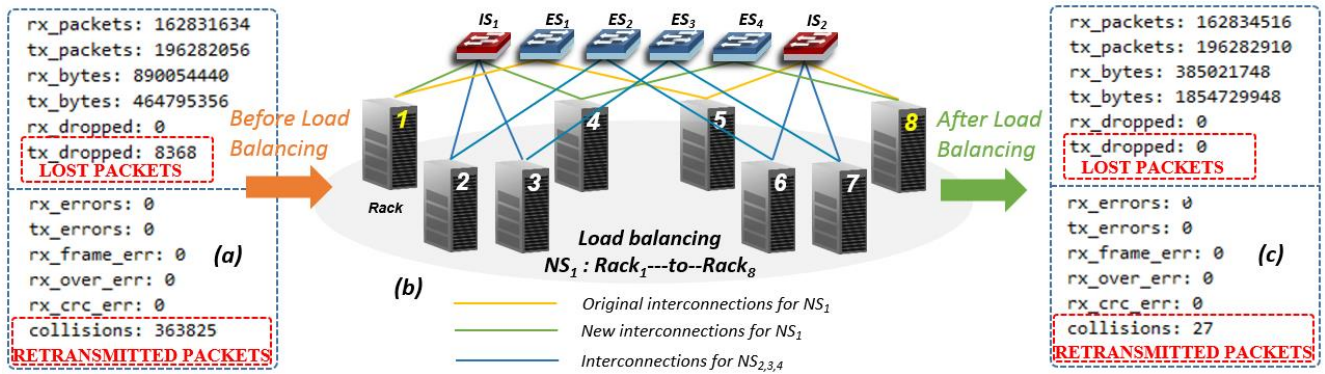

Fig. 8 Monitored statistics automatically trigger the load balancing.

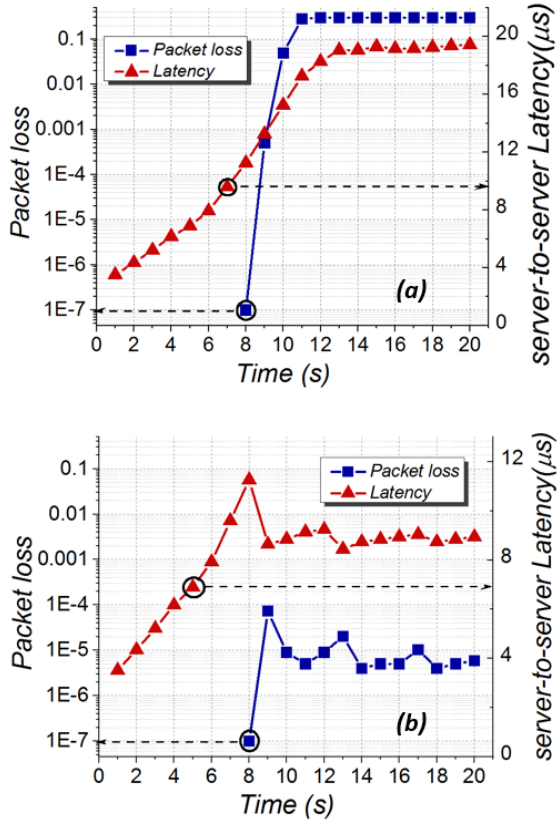

Fig. 9. Packet loss and latency performance without adjusting (a) and with load balancing (b).

### C. Statistics monitoring automatically trigger load balancing

The control plane is also in charge of monitoring and collecting the statistics of the underlying data plane. As the load increases, the contentions at switch nodes would cause high packet loss. Based on the collected real-time information, the data plane enables the automatic adjustments of the link interconnections for loading balancing to ensure the required QoS performance. To this end, the OF StatsReq – StatsRep commands are utilized between the ODL controller and interconnected hardware, to report the statistics information to the control plane. In particular in this experiment, the counts of the lost packet, which affects the QoS significantly, due to overfill of the buffer are collected at FPGA-based ToRs. Meanwhile the FPGA switch controllers calculate the number of retransmission packets (packets retransmission causes an extra latency) that is essentially the amounts of NACKs signals. This monitored information of $NS_1$ as shown in Fig. 8(a) is reported to the corresponding OF Agents and then then reported to the OS Orchestrator through the ODL controller. Once the Monitoring Engine detects that the packets losses at ToRs (the count of lost packets is 8368 in this case) are over a preventive packet loss threshold of 1E-5, Monitoring Engine automatically triggers the load balancing algorithm which generates an alternative path, to maintain the requested QoS. Triggered by the Monitoring Engine, the Optical Provisioning Manager coordinates with the PCE to calculate the new optical connectivity ($ToR_1 \leftrightarrow IS_1 \leftrightarrow ToR_4 \leftrightarrow ES_4 \leftrightarrow ToR_8$) between Rack-1 and Rack-8 for $NS_1$. Once the new network connections are deployed, the packet loss will be below the threshold and the retransmitted packets are decreased as well as shown in Fig. 8(c). We have measured the time evolution of the aggregated packet loss and latency for $NS_1$ without load balancing as shown in Fig. 9(a). The experiments have been repeated 5 times and the repeatable results are similar with less than ±3.7% deviations. If control plane does not update any of the LUTs to support the new NS interconnection, packet loss and latency keep increasing due to accumulated packet contentions. In comparison, Fig. 9(b) shows the network performance of $NS_1$ after load balancing. Targeting a packet loss threshold of 1E-5, once the reported statistics tend to exceed this value, the traffic load on original interconnection of $NS_1$ will be partly balanced to new connections implemented through the OF FlowMod commands. Therefore, packet loss below 1E-5 and server-to-server latency lower than 11.8 μs are achieved for high traffic load with load balancing operation.

## IV. CONCLUSION

An SDN-controlled and orchestrated OPSquare DCN enabling automatic NS configurations with dynamic QoS provisioning for multi-tenant applications has been experimentally investigated. We have experimentally validated the distributed and parallel data plane switching system deployed Optical Flow Control protocol to prevent packet loss at optical switch nodes. At the control plane, the OF protocol is properly extended to support the optical switching characteristics of the OPSquare. OF Agents are employed upon the connected switching nodes to facilitate the information exchange between the control and data planes. The experimental assessments demonstrate that automatic NSs provisioning and reconfiguration are implemented to satisfy dynamic QoS requirement of each NS, based on the collected resource topology. Moreover, the dynamic NS deployment significantly improves infrastructure utilization efficiency. Based on the functional abstractions of the data plane, the SDN control plane decouples the monitor and control operations from the underlying hardware without interfering the fast switching capability (nanoseconds) of optical switches. In addition, for NSs running applications with various latency requirements, flow priority can be dynamically assigned in the label field to prevent the packet retransmission caused latency performance degradation. Furthermore, the control plane is able to monitor the network performance by collecting real-time data plane statistics. The monitored statistics above the preventive threshold can automatically trigger the interconnections adjusting of NS. Therefore, the load balancing operations further enhance the QoS achieving packet loss below 1E-5 and server-to-server latency lower than 11.8 μs network performance even at a load of 1.0. Overall, with the implementation of an SDN-controlled and orchestrated control plane, the optical OPSquare DCN can be automatically configured to provide dynamic NSs deployment with specialized QoS provisioning and high resource utilization.